\documentclass[12pt]{article}
\usepackage{ifpdf}
\usepackage {graphics}
\usepackage{amsmath,mathptmx,amssymb,bm}%,amssymb,amsthm,enumerate,cite,mathrsfs%,backref

\newtheorem{thm}{Theorem}

\begin{document}

%\begin{frontmatter}

\title{Conservation laws for a generalized seventh order KdV equation}

%\author{ \\ Departamento de Matem\'aticas, Universidad de C\'adiz, 11510,
% \\ Puerto Real, C\'adiz, Spain\\\date{ {e.mail: 
% }}}
%\date{\empty}

\author{M.S. Bruz\'{o}n, A.P. M\'arquez, T.M. Garrido, E. Recio, R. de la Rosa\\
Universidad de C\'adiz, Spain\\
e-mail: m.bruzon@uca.es,almudena.marquez@uca.es,tamara.garrido@uca.es,\\
elena.recio@uca.es,rafael.delarosa@uca.es
}

%\author{University of C\'adiz}

%\author{R. de la Rosa${}^{a}$, M.L. Gandarias${}^{b}$, M.S. Bruz\'on${}^{c}$\\
% ${}^{a}$ Universidad de C\'adiz, Spain  (e-mail: rafael.delarosa@uca.es). \\
% ${}^{b}$ Universidad de C\'adiz, Spain (e-mail:  marialuz.gandarias@uca.es). \\
% ${}^{c}$ Universidad de C\'adiz, Spain  (e-mail:  m.bruzon@uca.es). \\  
%}

\date{}
 
\maketitle

\begin{abstract}

In this paper, by applying the multiplier method we obtain a complete classification of low-order local conservation laws for a generalized seventh-order KdV equation depending on seven arbitrary nonzero parameters. We apply the Lie method in order to classify all point symmetries admitted by the equation in terms of the arbitrary parameters. We find that there are no special cases of the parameters for which the equation admits extra symmetries, other than those that can be found by inspection (scaling symmetry and space and time translation symmetries). We consider the reduced ordinary differential equations and we determined all integrating factors of the reduced equation from the combined $x$- and $t$- translation symmetries. Finally, we observe that all integrating factors arise by reduction of the low-order multipliers of the generalized seventh-order KdV equation.\\

\noindent \textit{Keywords}: Partial differential equations, multiplier method, conservation laws, first integrals.\\
%\noindent \textit{PACS}: 02.20.Hj; 02.20.Sv; 02.30.Em; 02.30.Jr; 47.10.ab.\\
%\noindent \textit{MSC}: 35C07; 35Q53.
\end{abstract}

%\MOS{XXX; XXX}

%\keywords{Lie Symmetries; Nonlinear self-adjointness; Conservation
%laws; Multipliers}

\maketitle

\section{Introduction}

In this paper we study the following generalized seventh-order KdV equation:\begin{equation}\label{ed1}
\begin{array}{rl}
u_t+{a_1}\,{u}^{3}u_x+{a_2}\,u_x^{3}+{a_3}\,uu_{x}u_{2x}+{a_4}\,{u}^{2}u_{3x} +{a_5}\,u_{2x}u_{3x}& \\
+{a_6}\,u_xu_{4x}+{a_7}\,uu_{5x}+u_{7x}&=0,
\end{array}
\end{equation} where $a_i$ ($i = 1,\ldots, 7$) are nonzero parameters and $u_{nx}=\frac{\partial^n u}{\partial x^n}$ with $n=2,\ldots,7$.

The Korteweg-de Vries equation (KdV) was proposed to describe the steady translation water waves in a channel. The main property of this equation is the equal occurrence of non-linearity and dispersion. Some of its most important solutions, the solitary waves, are single, localised and non-dispersive entities, having also localised finite energy density. Several methods have been applied to solve the KdV equation such as the inverse scattering method or theoretical
group methods.

Many papers have been published recently studying KdV equations of different order \cite{kdv1,kdv2,kdv3,kdv4}. The integrability of this equation is important because this nonlinear evolution equation describes the behaviour of physical phenomena such as shallow water waves and plasmas. Therefore, its conservation laws are determined to predict its complete integrability.

For instance, in \cite{main} the author proved that the parametrized seventh-order KdV equation  passes the Painlev\'e test only in three distinct cases. Firstly, he made use of the truncated Painlev\'e expansion to derive their bilinear forms and then he constructed their multi-soliton solutions. Secondly, the pseudopotentials and Lax pairs were also derived. Finally, the infinite conservation laws were found using its Lax pair, and the author claimed that ``all conserved densities and fluxes'' are obtained with explicit recursion formulas. Moreover, in \cite{kdv1} the author obtained a soliton solution of the fifth-order KdV equation applying the homotopy analysis method (HAM). Besides, in \cite{kdv4} multiple-soliton solutions were obtained for the ninth-order KdV equation applying the Hirota bilinear method.

It is straightforward to see that equation (\ref{ed1}) is invariant under the scaling point symmetry $$
x \rightarrow  \lambda x,\quad
t \rightarrow  \lambda^7 t,\quad
u \rightarrow  \lambda^{-2} u.
$$
Note that equation (\ref{ed1}) contains all of the possible monomial terms with positive powers
of $u$ and derivatives of $u$ that preserve the scaling symmetry.
From \cite{BCA}  one can often find fluxes of conservation laws by an algebraic formula in terms  of the corresponding sets of multipliers without integration.

Conservation laws play a significant role in the solution process of an equation. They are very useful in the study of partial differential equations (PDEs) in which certain physical properties do not change in the course of time, especially for determining conserved quantities and constants of motion. Furthermore, they are good predicting linearizations, as well as checking the accuracy of numerical solution methods. Anco and Bluman proposed the multiplier method \cite{anco0,anco1a,anco1b} that gives a general treatment of a direct conservation law for PDEs. Some examples of the multiplier method can be found in \cite{anco2,anco3,ancoibra,ancon1,HHV,KH,MUA,RA,RIT}. In the case of evolution equations, there exists a one-to-one relationship between non-zero multipliers and non-trivial conserved currents (up to local equivalence). Moreover, conservation laws which possess a physical importance usually come from low-order multipliers. In addition, higher-order multipliers are related to the integrability of the equation.

In this work we consider the generalized seventh-order KdV equation (\ref{ed1}), we apply the multiplier method and we determine the parameters $a_i\neq 0$ with $i=1,\ldots 7$ for which equation (\ref{ed1}) admits multipliers. For each multiplier, we construct the conservation laws of equation (\ref{ed1}). Next, we perform an analysis of Lie symmetries and we use the symmetries to reduce the generalized seventh-order KdV (1) to an ODE. Furthermore, we find the integrating factors of the reduced ODE.

\section{Multipliers  and conservation laws of equation (\ref{ed1})}
Conservation laws are of basic importance because they provide physical, conserved quantities for all solutions $u(x, t)$,
and they can be used to check the accuracy of numerical solution methods. A general discussion of conservation laws
and their applications to differential equations can be found in \cite{ancorev,anco0,anco1a,BCA,HHV}.

A local conservation law for the generalized seventh-order KdV equation (\ref{ed1}) is a continuity equation
\begin{equation}\label{conslaw}
\displaystyle{D_t T+D_x X=0,}
\end{equation}
which is hold on all solutions of equation \eqref{ed1},
whereas the conserved density $T$ and the spatial flux $X$ are functions of $t$, $x$, $u$ and derivatives of $u$.
$D_t$ and $D_x$ denote the total derivative operators with respect to $t$ and $x$ respectively.
In the case that,
there is a function $\Theta(t,x,u,u_t,u_x,\dots)$ so that the conserved current vector $(T,X)=\left( D_x \Theta, -D_t \Theta \right)$ holds for every solution $u(t,x)$,
equation \eqref{conslaw} becomes an identity.
Thus, this conservation law is called locally trivial.
Thereby,
if two conservation laws differ by a locally trivial conservation laws, they are considered to be locally equivalent. A non-trivial conservation law can be written in a general form as follows
\begin{equation}
\frac{d}{dt} \left. \int_{\Omega} T { dx}=
-X \right|_{\partial \Omega},
\end{equation}
where $\Omega \subseteq \mathbb{R}$ is any fixed spatial domain. Any local conservation law can be stated by using the characteristic form which arises from a divergence identity
\begin{eqnarray}
\label{charcl} \nonumber
D_t \tilde{T}+D_x \tilde{X}
&=&(u_t+{a_1}\,{u}^{3}u_x+{a_2}\,u_x^{3}+{a_3}\,uu_{x}u_{2x}+{a_4}\,{u}^{2}u_{3x}\\&& +{a_5}\,u_{2x}u_{3x}
+{a_6}\,u_xu_{4x}+{a_7}\,uu_{5x}+u_{7x})Q,
\end{eqnarray}
where $\tilde{T}=T+D_x \Theta$ and $\tilde{X}=X-D_t \Theta$ are locally equivalent to $T$ and $X$, where the function $Q(t,x,u,u_t,u_x,\dots)$ which holds
\begin{equation}\label{eul}
Q= E_u (\tilde{T}),
\end{equation}
is called a multiplier,
and where $E_u$ represents the Euler operator with respect to $u$ \cite{olver},
\begin{equation}
E_u=
\partial_u-D_x \partial_{u_x}-D_t \partial_{u_t}+D_x D_t\partial_{u_{xt}}+D_x^2\partial_{u_{2x}}+.... \end{equation}
For evolution equations, there is a one-to-one correspondence between non-zero multipliers and non-trivial conserved current vectors up to local equivalence \cite{anco1a,olver}. This correspondence comes from the characteristic equation for multipliers of the given equation. Moreover, in the case of dispersive nonlinear water waves equations, conservation laws of physical importance typically come from low-order multipliers \cite{ancorev}.
The general form of low-order multipliers $Q$
in terms of $u$ and derivatives of $u$
is given by those variables that can be differentiated
to obtain a leading derivative of the KdV equation \eqref{ed1}.
Therefore,
the general form for a low-order multiplier
for equation \eqref{ed1} is
\begin{equation}
\label{mult}
Q(t,x,u,u_x,u_{2x},u_{3x},u_{4x},u_{5x},u_{6x}).
\end{equation}

Given a multiplier $Q$, we can obtain the conserved density using a standard method \cite{wolf}
\begin{equation}\label{dens} T= \int_{0}^{1}   \, u Q (x,t,\lambda u,\lambda u_x,\lambda u_{2x},...)\,d \lambda.\end{equation}

All low-order multipliers can be found by solving
the determining equation $E_u((u_t+{a_1} {u}^{3}u_x+{a_2} u_x^{3}+{a_3} uu_{x}u_{2x}+{a_4} {u}^{2}u_{3x} +{a_5} u_{2x}u_{3x}
+{a_6} u_xu_{4x}+{a_7} uu_{5x}+u_{7x})Q)=0$. In \cite{anco1a} from \eqref{eul} and \eqref{dens} it is defined the order of a conservation law for a generalized KdV equation as the order of the highest $x$ derivative of $u$ in its multiplier.

In order to obtain a general classification of low-order multipliers  \eqref{mult}, we write and split the determining equation \eqref{charcl} yielding an overdetermined system for  $Q$ and the parameters $a_i$ with $i=1,\ldots,7$. The solutions with $a_i\neq 0$ are given by

\begin{enumerate}
\item For $a_2=-a_4+\tfrac{1}{2}a_3$ and $a_3,a_4,a_5,a_6,a_7$ nonzero constants, 
\begin{equation}\label{Q1}Q_1 = 1\end{equation}

\item For $a_1={\tfrac {5}{49}} {a_6}^{3}-\tfrac{1}{49} {a_6}^{2}a_5
-{\tfrac {30 }{49}}{a_6}^{2}a_7+{\tfrac {3 }{49}}a_6 {a_5
} a_7+{\tfrac {225 }{196}}a_6 {a_7}^{2}-{\tfrac
{9 }{196}}a_5 {a_7}^{2}-{\tfrac {135}{196}} {a_7}^{3}
+\tfrac{3}{7} a_6 a_4+ \tfrac{2}{21} a_6 a_2-{\tfrac {9 }{14}}a_4
 a_7-\tfrac{1}{7} a_2 a_7,\quad a_3=-\tfrac{5}{7} {a_6}^{2}
+\tfrac{1}{7} a_5 a_6+{\tfrac {45 }{14}}a_6 a_7-\tfrac{3}{14} {
 a_5} a_7-{\tfrac {45 }{14}}{a_7}^{2}+a_4$, 
\begin{equation}\label{Q2}Q_2=\tfrac{1}{14}\left( 2 a_6-3 a_7 \right) {u}^{2}+ u_{2x}\end{equation}

%\item For $a_1=-{\tfrac {3a_5{a_7}^{2}}{98}}+{\tfrac {2}{49}}{
%a_5}a_7a_6-{\tfrac {2}{147
%}}a_5{a_6}^{2}-{\tfrac {45}{98}}{a_7}^{3}+{\tfrac {75
%}{98}}{a_7}^{2}a_6-{\tfrac {20}{49}}a_7{a_6}^{2}+{\tfrac {10}{147}}{{ a_6
%}}^{3}-\tfrac{4}{7}a_4a_7+{\tfrac {8}{
%21}}a_4a_6,\quad a_2=-{\tfrac {3a_5a_7}{28}}+\tfrac{1}{14}a_5{
% a_6}-{\tfrac {45}{28}}{a_7}^{2}+{\tfrac {45}{28}}a_7{ a_6
%}-{\tfrac {5}{14}}{a_6}^{2}-\tfrac{1}{2}a_4,\quad a_3=-\tfrac{3}{14}{
% a_5}a_7+\tfrac{1}{7}a_5a_6-{\tfrac {45}{
%14}}{a_7}^{2}+{\tfrac {45}{14}}a_7a_6-\tfrac {5}{7}{a_6}^{2}+a_4$ we find
%$Q_1$ and $Q_2$.

\item For $a_2=-3 a_4+a_3,\quad a_5=5 a_6-10 a_7$,  
\begin{equation}\label{Q3}Q_3=u\end{equation}

%\item For $a_2=a_4,a_3=4a_4,a_5=-10a_7+5 a_6$ besides $Q_1$, we find $Q_3$.

%\item For $a_1=-{\tfrac {15}{196}}{a_7}^{3}+{\tfrac {5}{49}}{a_7}^{2}
%a_6-{\tfrac {5}{147}}a_7{a_6}^{2}-{\tfrac {5
%}{14}} a_4a_7+{\tfrac {5}{21}}a_4a_6,\quad a_2=-
%2a_4-{\tfrac {15}{14}}{a_7}^{2}+\tfrac{5}{7}a_7a_6,\quad a_3=-{\tfrac {15}{14}}{a_7}^{2}+\tfrac{5}{7}a_7a_6+
%a_4,\quad a_5=-10a_7+5a_6$ we get $Q_2$ and $Q_3$.

%\item $a_5=-10a_7+5a_6,\quad a_1={\tfrac {5 }{882}}\left( 2a_6-3a_7 \right) ^{2}a_7,\quad a_2={\tfrac {5}{21}}a_7a_6-{\tfrac {5}{14}}{a_7}^{2},\quad a_4={
%\tfrac {5}{21}}a_7a_6-{\tfrac {5}{14}}{a_7}^{2},\quad a_3={\tfrac {20}{21}}a_7a_6-{\tfrac {10}{7}}{a_7}^{2
%}$ we obtain $Q_1$, $Q_2$ and $Q_3$.

\item For $a_1=-{\tfrac { 1}{147}}\left( a_6-4 a_7 \right)  \left( 4 {a_6}^{2}-18 a_6 a_7+8 {a_7}^{2}+49 a_4 \right) ,\quad a_2=\tfrac{1}{7} {a_6}^{2}-{\tfrac {9 }{14}}a_6 {a_7}+\tfrac{2}{7} {a_7}^{2}+2 a_4, \quad a_3=\tfrac{2}{7} {a_6}^{2}-{\tfrac {9 }{7}}a_6 a_7+\tfrac{4}{7} {a_7}^{2}+6 { a_4},\quad a_5=2 a_6-a_7$,

\begin{equation}
\begin{array}{ll}\label{Q4}
Q_4=& \left( {\tfrac {4 }{147}}{a_6}^{2}+{\tfrac {8}{147}} {a_7}^{2}+\tfrac{1}{3}a_4-{\tfrac {6}{49}} a_6 a_7 \right) {u}^{3}\\
 & + \left( \tfrac{2}{7} a_6-\tfrac{a_7}{7} \right) u_{2x}u+ \left( \tfrac{a_6}{7}-\tfrac{a_7}{14} \right) {u_{x}}^{2}+u_{4x},
\end{array}
\end{equation}

\begin{equation}
\begin{array}{ll}\label{Q5}
Q_5=& \left( {\tfrac { \left( 4 {a_6}^{2}-18 a_6 a_7+8 {{ a_7}}^{2}+49 a_4 \right)  \left( a_6-4 a_7 \right) {u_{{}}}^{3}}{147}} \right.\\
 & +\tfrac{1}{7}  \left( 2 a_6-a_7 \right)  \left( a_6-4 a_7 \right) u_{2x}u \\
 & +\tfrac{1}{14} \left( 2 a_6-a_7 \right)  \left( a_6-4 a_7 \right) {u_{x}}^{2}\left.+ \left( a_6-4 a_7 \right) u_{4x} \Big) t+x \right.
\end{array}
\end{equation}

\item For $a_2=a_4,a_3=4\,a_4,a_5=5\,a_7,a_6=3\,a_7$, 
\begin{equation}\label{Q6}
\begin{array}{ll}
Q_6=& \tfrac{1}{4}\,a_1\,{u_{{}}}^{4}+a_4\,{u_{{}}}^{2}u_{2x}+ \left( a_4\,{u_{{2}}}^{2}+a_7\,u_{4x} \right) u_{{}}+u_{6x}\\
& +2\,u_{x}a_7\,u_{3x}+3/2\,{u_{2x}}^{2}a_7
\end{array}
\end{equation}

\item For $a_1={\tfrac {5\,{a_7}^{3}}{98}},a_2={\tfrac {5\,{a_7}^{2}}{14}},a_3={\tfrac {10\,{a_7}^{2}}{7}},a_4=
{\tfrac{5\,{a_7}^{2}}{14}},a_5=5\,a_7,a_6=3\,a_7
$,
\begin{equation}\label{Q7}
\begin{array}{ll}
Q_7=&{\tfrac {5\,{a_7}^{2}{u}^{4}}{392}}+{\tfrac {5\,a_7\,{u_{{}}}^{2}u_{2x}}{14}}+ \left( u_{4x}+
{\tfrac {5\,a_7\,{u_{x}}^{2}}{14}} \right) u+{\tfrac {u_{6x}}{a_7}}+2\,u_{x}u_{3x}\\
 & +\tfrac{3}{2}\,{u_{2x}}^{2}
\end{array}
\end{equation}
 \end{enumerate}
Thus, we obtain the following result:
\begin{thm}
All nontrivial low-order local conservation laws admitted by the seventh-order KdV equation \eqref{ed1} with $a_i\neq 0$, $i=1,...,7$ are given by:

\begin{enumerate}
\item For $a_2=-a_4+\tfrac{1}{2}a_3$ and $a_3,a_4,a_5,a_6,a_7$ nonzero constants,
\begin{equation}
\begin{aligned}
&
T_1 = u
,\\ &
X_1 = u_{6x}+a_7 uu_{4x}+ \left( a_6-{a_7} \right) u_{x}u_{3x}
+ \left( \tfrac{1}{2}a_5-\tfrac{1}{2}a_6+\tfrac{1}{2}a_7 \right) {u_{2x}}^{2}\\ & \qquad+a_4 {u}^{2}u_{2x}
+ \left( -{ a_4}+\tfrac{1}{2}a_3 \right) u{u_{x}}^{2}+\tfrac{1}{4} a_1 {u}^{4}
\end{aligned}
\end{equation}

\item For $a_1={\tfrac {5}{49}} {a_6}^{3}-\tfrac{1}{49} {a_6}^{2}a_5
-{\tfrac {30 }{49}}{a_6}^{2}a_7+{\tfrac {3 }{49}}a_6 {a_5
} a_7+{\tfrac {225 }{196}}a_6 {a_7}^{2}-{\tfrac
{9 }{196}}a_5 {a_7}^{2}-{\tfrac {135}{196}} {a_7}^{3}
+\tfrac{3}{7} a_6 a_4+ \tfrac{2}{21} a_6 a_2-{\tfrac {9 }{14}}a_4
 a_7-\tfrac{1}{7} a_2 a_7,\quad a_3=-\tfrac{5}{7} {a_6}^{2}
+\tfrac{1}{7} a_5 a_6+{\tfrac {45 }{14}}a_6 a_7-\tfrac{3}{14} {
 a_5} a_7-{\tfrac {45 }{14}}{a_7}^{2}+a_4$,
\begin{equation}
\begin{aligned}
&
T_2 =
\tfrac{1}{3} A {u}^{3}-7 {u_{x}}
^{2}
,\\ &
X_2 =
 \left(  A {u}^{2}+14 u_{2x} \right) u_{6x}+7 {u_{4x}}^{2}+ \left( -14 u_{3x} +\left( -4 {a_6}+6 a_7 \right) uu_{x} \right) u_{5x}\\ & \qquad
 + \left(  \left( 4 a_6+8 a_7 \right) uu_{2x}
 + \left( 4 a_6-6 a_7 \right) {u_{x}}^{2} + a_7 A {u}^{3} \right)u_{4x}
\\ & \qquad
+ \left(  \left( 2 a_6+4 a_7 \right) u_{x}u_{2x}+ A  \left( a_6-3 {a_7} \right) {u}^{2}u_{x} \right) u_{3x}
\\ & \qquad
+ \left( \tfrac{14}{3} {a_5}-\tfrac{2}{3} a_6-\tfrac{4}{3} a_7 \right) {u_{2x}}^{3}+14 u_{x}u_{x}+ \left( -2 a_6-4 a_7 \right) u{u_{3x}}^{2} \\ & \qquad
 + \left( -\tfrac{3}{2} a_5 a_7+\tfrac{9}{2} a_6 a_7-{a_6}^{2}-\tfrac{9}{2} {{a_7}}^{2}+7 a_4+a_5 a_6 \right) {u}^{2}{u_{2x}}^{2} \\ & \qquad
+ \left( -2  A  \left(a_6-3 a_7 \right) u{u_{x}}^{2}+a_4  \left( 2a_6-3 a_7 \right) {u}^{4} \right) u_{2x} \\ & \qquad
+ \left( {{ a_6}}^{2}-\tfrac{9}{2} a_6 a_7+\tfrac{9}{2} {a_7}^{2}+\tfrac{7}{2} a_2 \right) {u_{x}}^{4}+\tfrac{1}{3} a_2  \left( 2 a_6-3 a_7
 \right) {u}^{3}{u_{x}}^{2}  \\ & \qquad
+{\tfrac {1}{3528}} A^{2}\left( -6 a_5 a_6+9
a_5 a_7+30 {a_6}^{2}-135 a_6 a_7+135 {{ a_7}}^{2}+28 a_2+126 a_4 \right) {u}^{6}
\end{aligned}
\end{equation}
where $A=2 a_6-3 a_7 $
\item For $a_2=-3 a_4+a_3,\quad a_5=5 a_6-10 a_7$,
 \begin{equation}
\begin{aligned}
&
T_3 =\tfrac{1}{2} {u}^{2}
,\\ &
X_3 =uu_{6x}-u_{5x}u_{x}+ \left( a_7 {u}^{2}+u_{2x} \right) u_{4x}-\tfrac{1}{2} {u_{3x}}^{2}\\
 & \qquad + \left( a_6-2 a_7 \right) uu_{x}u_{3x} + \left(2 a_6-4 a_7 \right) u{u_{2x}}^{2}
\\
& \qquad+ \left(  \left(-a_6+2 a_7 \right) {u_{x}}^{2}+a_4 {u}^{3} \right) u_{2x}+ \left( \tfrac{1}{2}a_3-\tfrac{3}{2} a_4 \right) {u}^{2}{u_{x}}^{2}\\
& \qquad +\tfrac{1}{5} a_1 {u}^{5}
\end{aligned}
\end{equation}

\item For $a_1=-{\tfrac { 1}{147}}\left( a_6-4 a_7 \right)  \left( 4 {a_6}^{2}-18 a_6 a_7+8 {a_7}^{2}+49 a_4 \right) ,\quad a_2=\tfrac{1}{7} {a_6}^{2}-{\tfrac {9 }{14}}a_6 {a_7}+\tfrac{2}{7} {a_7}^{2}+2 a_4, \quad a_3=\tfrac{2}{7} {a_6}^{2}-{\tfrac {9 }{7}}a_6 a_7+\tfrac{4}{7} {a_7}^{2}+6 { a_4},\quad a_5=2 a_6-a_7$

\begin{equation}
\begin{aligned}
&
T_4 =\tfrac{1}{2} {u_{2x}}^{2}+ \tfrac{1}{14}\left( a_7-2a_6 \right) u{u_{x}}^{2}\\
 & \qquad + \left( {\tfrac {1}{147}}{a_6}^{2}-{\tfrac {3 a_7a_6}{98}}+{\tfrac {2}{147}} {a_7}^{2}+\tfrac{1}{12}a_4 \right) {u}^{4}
,%\\ &
\end{aligned}
\end{equation}
%\begin{equation}\label{T_4}
%T_4=1/2 {u_{2x}}^{2}+ \left( a_7/14-a_6/7 \right) u{u_{
%{2}}}^{2}+ \left( {\tfrac {{a_6}^{2}}{147}}-{\tfrac {3 a_7 {
%\it a6}}{98}}+{\tfrac {2 {a_7}^{2}}{147}}+a_4/12 \right) {u_
%{{}}}^{4}
%\end{equation} and $X_4$
\item For $a_1=-{\tfrac { 1}{147}}\left( a_6-4 a_7 \right)  \left( 4 {a_6}^{2}-18 a_6 a_7+8 {a_7}^{2}+49 a_4 \right) ,\quad a_2=\tfrac{1}{7} {a_6}^{2}-{\tfrac {9 }{14}}a_6 {a_7}+\tfrac{2}{7} {a_7}^{2}+2 a_4, \quad a_3=\tfrac{2}{7} {a_6}^{2}-{\tfrac {9 }{7}}a_6 a_7+\tfrac{4}{7} {a_7}^{2}+6 { a_4},\quad a_5=2 a_6-a_7$
\begin{equation}
\begin{aligned}
&
T_5 = \left( \tfrac{a_6}{2}-2 a_7 \right) t{u_{2x}}^{2}-\tfrac{1}{14}  \left(
-a_7+2 a_6 \right)  \left( a_6-4 a_7 \right) tu_
{{}}{u_{x}}^{2}\\&\qquad
+{\tfrac { \left( 4 {a_6}^{2}-18 a_7 {
 a_6}+8 {a_7}^{2}+49 a_4 \right)  \left( a_6-4 {
a_7} \right) t{u}^{4}}{588}}+xu
,%\\ &
\end{aligned}
\end{equation}

\item For $a_2=a_4,a_3=4\,a_4,a_5=5\,a_7,a_6=3\,a_7$
\begin{equation}
\begin{aligned}
&
T_6 =-\tfrac{1}{2} {u_{3x}}^{2}+\tfrac{1}{2} a_7 u{u_{2x}}^{2}-\tfrac{1}{2} { a_4} {u}^{2}{u_{x}}^{2}+\tfrac{1}{20} a_1 {u}^{5}
,%\\ &
%X_6 =\tfrac{1}{2} {u_{6x}}^{2}
%\\& \qquad
%+ \left( a_7u u_{4x}+2u_{x}a_7 u_{3x}+\tfrac{3}{2} {u_{2x}}^{2}a_7+a_4 {u}^{2}u_{2x}
%\right.\\&\qquad\left.
%+a_4 u{u_{x}}^{2}+\tfrac{1}{4} a_1 u_{}^{4} \right) u_{6x}
%\\& \qquad
%+u_{{t}}u_{5x}+\tfrac{1}{2} {u}^{2}{a_7}^{2}{u_{4x}}^{2}
%\\& \qquad
%+ \left( -u_{tx}+a_4 {u}^{3}u_{2x}a_7+a_4 {u}^{2}{u_{x}}^{2}a_7
%+2 u_{x}{a_7}^{2}u_{3x}u
%\right.\\&\qquad\left.
%+\tfrac{1}{4} a_1 {u}^{5}a_7+\tfrac{3}{2} {u_{2x}}^{2}{a_7}^{2}u \right) u_{4x}
%\\& \qquad
%+2 {u_{x}}^{2}{a_7}^{2}{u_{3x}}^{2}+ \left( u_{t2x}+a_7 u u_{t}+3 {u_{2x}}^{2}{a_7}^{2}u_{x}
%\right.\\&\qquad\left.
%+2 a_4 {u}^{2}u_{2x}u_{x}a_7
%+\tfrac{1}{2} a_1 {u}^{4}u_{x}a_7+2 a_4 u{u_{x}}^{3}a_7 \right) u_{{3x}}
%\\& \qquad
%+{\tfrac {9}{8}} {u_{2x}}^{4}{a_7}^{2}+\tfrac{3}{2} a_4 {u}^{2}{u_{2x}}^{3}a_7
% \\& \qquad
% + \left( \tfrac{3}{2} a_4 u{u_{{x}}}^{2}a_7+ \left( \tfrac{1}{2} {a_4}^{2}+\tfrac{3}{8} a_1 a_7 \right) {u}^{4} \right) {u_{2x}}^{2}
%  \\& \qquad
% + \left( {a_4}^{2}{u}^{3}{u_{x}}^{2}-a_7u u_{{tx}}+a_7 u_{t}u_{{2}}+\tfrac{1}{4} a_1 {u}^{6}a_4 \right) u_{2x}
%   \\& \qquad
% +\tfrac{1}{32}{{a_1}}^{2}{u}^{8}+\tfrac{1}{2} {a_4}^{2}{u}^{2}{u_{x}}^{4}
%+\tfrac{1}{4} a_1 {u}^{5}a_4 {u_{x}}^{2}+a_4 {u}^{2}u_{{1}}u_{x}
\end{aligned}
\end{equation}
\item For $a_1={\tfrac {5\,{a_7}^{3}}{98}},a_2={\tfrac {5\,{a_7}^{2}}{14}},a_3={\tfrac {10\,{a_7}^{2}}{7}},a_4=
{\tfrac{5\,{a_7}^{2}}{14}},a_5=5\,a_7,a_6=3\,a_7
$
\begin{equation}
\begin{aligned}
&
T_7 =-\frac{1}{2} a_7 t{u_{2x}}^{2}+{\frac {5 {a_7}^{2}tu{u_{x}}^{2}}{14}}-{\frac {5 {a_7}^{3}t{u}^{4}}{392}}+xu,\\ &
%X_7 =
%\left(  \left( -{\frac {5 {a_7}^{3}{u}^{3}}{98}}-\frac{5}{7} {{
%a_7}}^{2}uu_{2x}-{\frac {5 {u_{x}}^{2}{a_7}^{2}}{14
%}}-a_7 u_{4x} \right) u_{6x}+\right.\\&\qquad
% \frac{1}{2} a_7 {u
% 	_{5x}}^{2}+\left( {\frac {10 u_{x}{a_7}^{2}u_{2x}
%}{7}}+\frac{5}{7} {a_7}^{2}uu_{3x}+{\frac {15 u_{x}{{a_7
%}}^{3}{u}^{2}}{98}} \right) u_{5x} \\&\qquad
%-\frac{6}{7} {a_7}^{2}{u_{4x}}^{2}u+\left( -{\frac {15 u_{x}{a_7}^{2}u_{3x}}{7}}-{\frac {85 {a_7}^{3}{u}^{2}u_{2x}}{98}}-\right.\\&\qquad\left.
%{
%	\frac {65 {u_{x}}^{2}{a_7}^{3}u}{98}}-{\frac {10 {{a_7}}^{2}{u_{2x}}^{2}}{7}}-{\frac {5 {a_7}^{4}{u}^{4}}{98
%}} \right) u_{4x}+{\frac {25 {a_7}^{3}{u}^{2}{u_{3x}}^{2}}{98}}+
%\\&\qquad
%\left( -{\frac {25 {a_7}^{3}uu_{2x}u_{x}}{49}}-{\frac {75 {u_{x}}^{3}{a_7}^{3}}{98}}-a_7 u_{t} \right) u_{3x}-{\frac {50 {a_7}^{3}u{u_{2x}}^{
%3}}{49}}+\\&\qquad
% \left( {\frac {25 {u_{x}}^{2}{a_7}^{3}}{49}}-{\frac
%{25 {a_7}^{4}{u}^{3}}{98}} \right) {u_{2x}}^{2}+ \left(
%-{\frac {25 {u_{x}}^{2}{a_7}^{4}{u}^{2}}{196}}+a_7
%u_{tx}-
%\right.\\&\qquad\left.
%{\frac {25 {a_7}^{5}{u}^{5}}{1372}} \right) u_{2x} -\frac{5}{7} {a_7}^{2}uu_{t}u_{x}
%-{\frac {25 {a_7}^{6
%}{u}^{7}}{67228}}-{\frac {25 {a_7}^{5}{u}^{4}{u_{x}}
%^{2}}{2744}}\\&\qquad\left.
% -{\frac {25 {u_{x}}^{4}{a_7}^{4}u}{196}}
% \right) t+ \left( {\frac {5 {a_7}^{2}u{u_{x}}^{2}}{14}}+
%\frac{3}{2} {u_{2x}}^{2}a_7+{\frac {5 {a_7}^{2}{u}^{2}u_{xx}}{14}}
%\right.\\&\qquad\left.
%+2 u_{x}a_7 u_{3x}+ua_7 u_{4x}+u_{6x}+{\frac {5 {a_7}^{3}{u}^{4}}{392}}
% \right) x\\&\qquad
% -u_{5x}-a_7 u_{x}u_{2x}-ua_7
%u_{3x}-{\frac {5 {a_7}^{2}{u}^{2}u_{x}}{14}}
\end{aligned}
\end{equation}
$X_4$, $X_5$, $X_6$ and $X_7$ are omitted here in order to save space.
  \end{enumerate}
\end{thm}

\section{Lie symmetries}

In the Lie method, we required that the one-parameter Lie group of infinitesimal point transformations in $(x,t,u)$ given by
$$\begin{array}{lll}
 x^*&=&x+\epsilon \xi(x,t,u)+O(\epsilon^2),\\
 t^*&=&t+\epsilon \tau(x,t,u)+O(\epsilon ^2),\\
 u^*&=&u+\epsilon \eta(x,t,u)+O(\epsilon ^2),
\end{array}$$
leaves invariant the set of solutions of equation (\ref{ed1}), where  $\epsilon$ is the group parameter.

Hence, an infinitesimal point symmetry for the generalized seventh-order KdV equation (\ref{ed1}) is a generator
\begin{equation}\label{ps}
X=\xi(x,t,u) \frac{\partial}{\partial x}+ \tau(x,t,u)\frac{\partial}{\partial t}+\eta(x,t,u)\frac{\partial}{\partial u},
\end{equation}
whose prolongation leaves invariant the equation (\ref{ed1}). Every point symmetry (\ref{ps}) can be expressed by its characteristic form
$$ \hat{X} = P \partial_u, \ P=\eta - \tau u_t - \xi u_x.$$
The condition for the invariance of the equation (\ref{ed1}) under a point symmetry (\ref{ed1}) is given by
\begin{equation}\label{cs}
\begin{array}{rl}
0= & \left( 3a_1 u^2 u_x + a_3 u_x u_{2x} + 2 a_4 u u_{3x} + a7 u_{5x} \right) P  + D_tP\\
 & + \left( a_1 u^3 + 3a_2 u_x^2 + a_3 u u_{2x} + a_6 u_{4x} \right) D_xP \\
 & + \left( a_3 u u_x + a_5 u_{3x} \right) D_x^2 P + \left( a_4 u^2 + a_5 u_{2x} \right) D_x^3 P \\
 & + a_6 u_x D_x^4 P + a_7 u D_x^5 P + D_x^7 P
\end{array}
\end{equation}
and holds for all solutions $u(x, t)$ of equation (\ref{ed1}). Equation (\ref{cs}) is known as the symmetry determining equation. It splits with respect to $x$ derivatives of $u$. This leads to a linear overdetermined system of equations on $\xi(x,t,u)$, $\tau(x,t,u)$, $\eta(x,t,u)$ along with $a_1, a_2, a_3, a_4, a_5, a_6, a_7$.

Solving the system, the only point symmetries admitted by (\ref{ed1}) with $a_i\neq 0$, $i=1,\ldots, 7$, which are defined by the infinitesimal, are generated
$$\begin{array}{l}
{\bf X}_1=\displaystyle\frac{\partial}{\partial x},\qquad {\bf X}_2=\displaystyle\frac{\partial}{\partial t},\qquad
{\bf X}_3=x\frac{\partial}{\partial x}+ 7 t \frac{\partial}{\partial t}-2 u \frac{\partial}{\partial u}.
\end{array}$$

\noindent From ${\bf X}_1+{\bf X}_2$, we obtain travelling wave reductions $$\begin{array}{lll}
z=x-\lambda t,&\hspace{0.5cm} &u=h(z),\end{array}$$
where $h(z)$ satisfies
\begin{eqnarray}\nonumber h_{7z}+{ a_7} h h_{5z}+{ a_6}
 h_{z} h_{4z}+{ a_5} h_{2z} h_{3z}+{ a_4} h^
 2 h_{3z}+{ a_3} h h_{z} h_{2z}&&\\\label{ecf}+{ a_2} \left(h_{z}
 \right)^3+{ a_1} h^3 h_{z}-{ \lambda} h_{z}
&=&0\end{eqnarray}
with $h_{nz}=h^{n)}(z)$ for $n=2,...,7$ (the $n$-th order derivative of $h(z)$ ).

\noindent From ${\bf X}_3$ we obtain the invariant solution
\begin{equation}\label{trans}\begin{array}{lll} z= \displaystyle\frac{x^7}{t},&\hspace{0.5cm}
&u=\displaystyle{{h}\over{t^{{{1}\over{4\,n+4}}}}}x^{-2}h(z),\end{array}\end{equation} where $h(z)$ must satisfy a non-autonomous equation.

In \cite{SK1,SK2}  used the complex singularity structure of solutions  to identify an integrable case of the equations of motion for a
rotating top.  Ablowitz and Segur \cite{AS}, and Ablowitz,
Ramani, and Segur \cite{ARS1,ARS2,ARS3} observed this connection in
the context of integrable PDEs. Their observations led to the following conjecture:

 {\it Any ODE which arises as a reduction
of an integrable PDE possesses the Painlev\'e property, possibly
after a transformation of variables}.

In \cite{main} the author showed that the parameterized seventh-order
KdV equation (\ref{ed1}) passes the Painlev\'e test only in three distinct
cases:

\begin{itemize}
\item $a_1=\frac{4}{147},\,a_2=\frac{1}{7},\,a_3=\frac{6}{7},\,a_4=\frac{2}{7},
    \,a_5=3,\,a_6=2,\,a_7=1$
\item $a_1=\frac{5}{98},\,a_2=\frac{5}{14},\,a_3=\frac{10}{7},\,a_4=\frac{5}{14},
    \,a_5=5,\,a_6=3,\,a_7=1$
\item $a_1=\frac{4}{147},\,a_2=\frac{5}{14},\,a_3=\frac{9}{7},\,a_4=\frac{2}{7},
    \,a_5=6,\,a_6=\frac{7}{2},\,a_7=1$
\end{itemize}
Consequently, by applying the conjecture, the reduced equations for these cases possess the Painlev\'e property.

\section{Integrating factors and first integrals for equation \eqref{ecf}} \label{dr}

In \cite{sjoberg}, Sj\"oberg proposed the so-called double reduction method which took into consideration the relationship between symmetries and conservation laws. This method can be used to reduce a $q$th-order PDE into a $(q-1)$th-order ODE. In \cite{a1}, by using the direct method of the multipliers \cite{anco1a,anco1b}, Anco developed symmetry properties of conservation laws of PDEs. In particular, the author provided a theoretical framework to comprehend and generalize the method proposed by Sj\"oberg. In this way, the author proved that conservation laws that are symmetry invariant or symmetry homogeneous have at least
one important application: any symmetry-invariant conservation
law will reduce to a first integral for the ODE obtained by symmetry reduction of the given PDE when symmetry-invariant solutions $u(t,x)$ are investigated. This provides a direct reduction of order of the ODE.

It is well known that for any $n$th-order ODE finding an integrating factor is equivalent to finding a first integral (see e.g. \cite{AB98,BA}). In this section, we focus our attention on obtaining integrating factors of ODE (\ref{ecf}) and, hence, first integrals. Furthermore, a first integral of equation (\ref{ecf}) yields a quadrature which reduces (\ref{ecf}) to a 6th-order ODE in terms of the original variables $z$, $h$, $h_z, \,\ldots, \, h_{6z}$.

Following reference \cite{BA} an integrating factor of equation (\ref{ecf}) is a function  $Q(z,h,h_z,\ldots,h_{lz})\neq 0$, with $0\leq l\leq 6$, which verifies the characteristic equation
\begin{eqnarray}\nonumber
D_z \tilde X&=& \left(h_{7z}+{ a_7} h h_{5z}+{ a_6}
 h_{z} h_{4z}+{ a_5} h_{2z} h_{3z}+{ a_4} h^
 2 h_{3z}+{ a_3} h h_{z} h_{2z}\right.\\\label{chareqn}&&\left.+{ a_2} \left(h_{z}
 \right)^3+{ a_1} h^3 h_{z}-{ \lambda} h_{z}\right)Q.
\end{eqnarray}

Moreover, all first integrals of ODE (\ref{ecf}) derive from integrating factors verifying the characteristic equation (\ref{chareqn}) for any function $h(z)$. The determining equation to obtain all integrating factors is
\begin{eqnarray}\nonumber
\frac{\delta}{\delta h}\Big( (h_{7z}+{ a_7} h h_{5z}+{ a_6}
 h_{z} h_{4z}+{ a_5} h_{2z} h_{3z}+{ a_4} h^
 2 h_{3z}+{ a_3} h h_{z} h_{2z}\\\label{multreqn}+{ a_2} \left(h_{z}
 \right)^3+{ a_1} h^3 h_{z}-{ \lambda} h_{z})Q \Big) &=&0,
\end{eqnarray}
where $\frac{\delta}{\delta h}$ represents the variational derivative. This equation must hold off of the set of solutions of equation (\ref{ecf}). Once the integrating factors are found,
the corresponding non-trivial first integrals are obtained by using a line integral \cite{BA}.\\

\noindent The determining equation  (\ref{multreqn}) splits with respect to the variables
$h_{7z}\ldots,h_{10z}$.
This yields a linear determining system for $Q$. Then, we determine all integrating factors:
\begin{enumerate}
\item For $a_2=-a_4+\frac{1}{2}a_3$ and $a_3,a_4,a_5.a_6,a_7$ nonzero constants
\begin{equation}\label{Q1}Q_1 = 1\end{equation}

\item For $a_1={\frac {5}{49}} {a_6}^{3}-\frac{1}{49} {a_6}^{2}a_5
-{\frac {30 }{49}}{a_6}^{2}a_7+{\frac {3 }{49}}a_6 {a_5
} a_7+{\frac {225 }{196}}a_6 {a_7}^{2}-{\frac
{9 }{196}}a_5 {a_7}^{2}-{\frac {135}{196}} {a_7}^{3}
+\frac{3}{7} a_6 a_4+ \frac{2}{21} a_6 a_2-{\frac {9 }{14}}a_4
 a_7-\frac{1}{7} a_2 a_7,\quad a_3=-\frac{5}{7} {a_6}^{2}
+\frac{1}{7} a_5 a_6+{\frac {45 }{14}}a_6 a_7-\frac{3}{14} {
 a_5} a_7-{\frac {45 }{14}}{a_7}^{2}+a_4$, we obtain the multiplier
\begin{equation}\label{Q2}Q_2=\frac{1}{14}\left( 2 a_6-3 a_7 \right) {h}^{2}+ h_{2z}\end{equation}

\item For $a_2=-3 a_4+a_3,\quad a_5=5 a_6-10 a_7$,
\begin{equation}\label{Q3}Q_3=h\end{equation}

\item For $a_1=-{\frac { 1}{147}}\left( a_6-4 a_7 \right)  \left( 4 {a_6}^{2}-18 a_6 a_7+8 {a_7}^{2}+49 a_4 \right) ,\quad a_2=\frac{1}{7} {a_6}^{2}-{\frac {9 }{14}}a_6 {a_7}+\frac{2}{7} {a_7}^{2}+2 a_4, \quad a_3=\frac{2}{7} {a_6}^{2}-{\frac {9 }{7}}a_6 a_7+\frac{4}{7} {a_7}^{2}+6 { a_4},\quad a_5=2 a_6-a_7$

\begin{eqnarray}\label{Q4}\nonumber
Q_4=& \left( {\frac {4 }{147}}{a_6}^{2}+{\frac {8}{147}} {a_7}^{2}+\frac{1}{3}a_4-{\frac {6}{49}} a_6 a_7 \right) {h}^{3}+ \left( \frac{2}{7} a_6-\frac{a_7}{7} \right) h_{2z}h\\
 & + \left( \frac{a_6}{7}-\frac{a_7}{14} \right) {h_{z}}^{2}+h_{4z},
\end{eqnarray}

\item For $a_2=a_4,a_3=4\,a_4,a_5=-10a_7+5a_6$
\begin{eqnarray}\label{Q5}Q_5=&\frac{1}{4}a_1 {h_{{}}}^{4}+a_4 {h}^{2}h_{2z}+ \left( a_4 {h_{z}}^{2}+a_7 h_{4z} \right) h_{{}}+h_{6z}+2 a_6h_{2z}^2\\&-\frac{9}{2} a_7 {h_{2z}}^{2}+a_6h_zh_{3z}-a_7h_zh_{3z}\end{eqnarray}
  \end{enumerate}
We observe that the integrating factors arise by reduction of the multipliers for equation \eqref{ed1}.

\section{Conclusions}\label{conclusions}

For a generalized seventh-order KdV equation we have derived  the low-order conservation laws by using
the multiplier method. Moreover  we have studied  equation \eqref{ed1} from Lie symmetries viewpoint and we consider the reduced ordinary differential equation under the combined space and time translation symmetries. We have determined all integrating factors of an ordinary differential equations and we observe that the  integrating factors arise by reduction of a conservation law for equation \eqref{ed1}.

%\newpage

\section*{Acknowledgements}

We gratefully acknowledge fruitful discussions and contributions received by Dr. Stephen Anco (Brock University). The authors express their sincere gratitude to the Plan Propio de Investigaci\'on de la Universidad de C\'adiz.

%%\printbibliography
%\bibliography{CMMSE_2018_BMGRdR-Rv-6}

\end{document}